\documentclass{INTERSPEECH2023}

\interspeechcameraready

\newcommand{\figref}[1]{Figure~\ref{#1}}
\newcommand{\tabref}[1]{Table~\ref{#1}}
\usepackage{amsmath}

\usepackage{multirow}

\title{Enhancing the EEG Speech Match Mismatch Tasks With Word Boundaries}
\name{Akshara Soman, Vidhi Sinha, Sriram Ganapathy}
\address{
  LEAP Lab, Indian Institute of Science Bangalore, India}
\email{aksharas@iisc.ac.in, vidhis@iisc.ac.in, sriramg@iisc.ac.in}

\begin{document}

\maketitle
 
\begin{abstract}
Recent studies have shown that the underlying neural mechanisms of human speech comprehension can be analyzed using a match-mismatch classification of the speech stimulus and the neural response. However, such studies have been conducted for fixed-duration segments without accounting for the discrete processing of speech in the brain. In this work, we establish that word boundary information plays a significant role in sentence processing by relating EEG to its speech input. We process the speech and the EEG signals using a network of convolution layers. Then, a word boundary-based average pooling is performed on the representations, and the inter-word context is incorporated using a recurrent layer.  The experiments show that the modelling accuracy can be significantly improved (match-mismatch classification accuracy) to 93\% on a publicly available speech-EEG data set, while previous efforts achieved an accuracy of 65-75\% for this task.



\end{abstract}

\section{Introduction}
Humans have the unique ability to communicate through speech. While speech comprehension is mastered from a young age, many neural processes enabling this seamless activity are unknown. One of the simplest  ways of furthering the understanding of speech comprehension is through the recording of neural responses  using electroencephalography (EEG). 
The EEG is a non-invasive neural imaging technique that measures electrical activity in the brain by placing electrodes on the scalp \cite{sanei2013eeg}.  It has been demonstrated that the EEG signal recorded during a speech listening task contains information about the stimulus \cite{gillis2021neural}. One can investigate how the brain comprehends continuous speech by developing models that relate the speech with  the EEG signal using machine learning techniques \cite{sanei2021eeg}. 

The early attempts explored linear models for relating continuous natural speech to EEG responses \cite{ding2012neural, crosse2016multivariate, wong2018comparison,vanthornhout2018speech}. They can be categorized into three different types - forward models, backward models, or hybrid models.   The forward models predict EEG from speech stimuli, while the backward models reconstruct speech from EEG responses. In many studies, the correlation between the predicted and ground truth signal is considered as a measure of neural tracking \cite{de2018decoding}.  However, linear models may be ill-equipped to capture the non-linear nature of the auditory system. Recently, deep neural networks have  been employed to compare and analyze speech stimuli and EEG responses. Several studies have shown promising results with deep learning models for EEG-speech decoding  \cite{monesi2020lstm,katthi2021deep,de2021auditory,accou2020modeling}.
These advancements  in speech decoding from the brain will also be beneficial for the development of brain-computer interfaces(BCIs). 

In many of the computational approaches, the speech envelope has been the most commonly used feature  \cite{ding2012neural,crosse2016multivariate,vanthornhout2018speech}. Other features such as spectrograms \cite{di2015low,lesenfants2019data}, phonemes \cite{di2015low}, linguistic  features \cite{lesenfants2019data,teoh2022attention}, and phono-tactics \cite{di2019low} have also been explored with linear forward/backward models. Lesenfants et al. ~\cite{lesenfants2019predicting} demonstrated that combining phonetic and spectrogram features improves the EEG-based speech reception threshold (SRT) prediction.

While forward/backward models and correlation tasks were previously explored, the match mismatch tasks have been recently investigated as an alternative task \cite{puffay2022relating, de2021auditory}. Here, the task is to identify whether a portion of the brain response (EEG) is related to the speech stimulus that evoked it. In  the previous studies using the match mismatch task, the auditory stimulus and speech of a fixed duration ($5$s) are processed through a series of convolutional and recurrent layers \cite{monesi2020lstm, accou2020modeling, puffay2023relating}.  

In this work, we argue that the prior works on speech-EEG match mismatch tasks are incomplete without considering the fragmented nature of speech comprehension. While speech and EEG signals are continuous, the neural tracking of speech signals is impacted by the linguistic markers of speech \cite{brodbeck2018rapid}. The most striking of this evidence comes from models of word surprisal \cite{koskinen2020brain} with N400 response evoked for unpredictable words \cite{kutas2011thirty, soman2022erp}. In the simplest form, we hypothesize that the task of relating continuous speech with EEG must also include word-level segmentation information.  

We propose a deep learning model to perform match mismatch classification tasks on variable length inputs using word boundary information. 
The model consists of convolutive feature encoders of both the speech and EEG inputs. Further, the word segmentation information, obtained by force-aligning the speech with the text data using a speech recognition system, is incorporated in the feature outputs through a word-level pooling operation. The pooled representations are further modelled with recurrent long short-term memory (LSTM) layers to model the inter-word context. The final output from the LSTM network for the speech and EEG streams is used in the match mismatch classification task.  

The major contributions of this paper are:
\begin{itemize}
    \item Proposing a match mismatch classification model that can incorporate word boundary information.
    \item Proposing a loss function based on Manhattan distance for the match mismatch task. 
    \item Experimental illustration of the effectiveness of the model, where the classification performance is significantly improved over the prior works. 
    \item A detailed set of ablation experiments to elicit the impact of word boundary information in speech EEG matching task. 
\end{itemize}

\begin{figure}[t!]
  \centering
  \includegraphics[width=\linewidth]{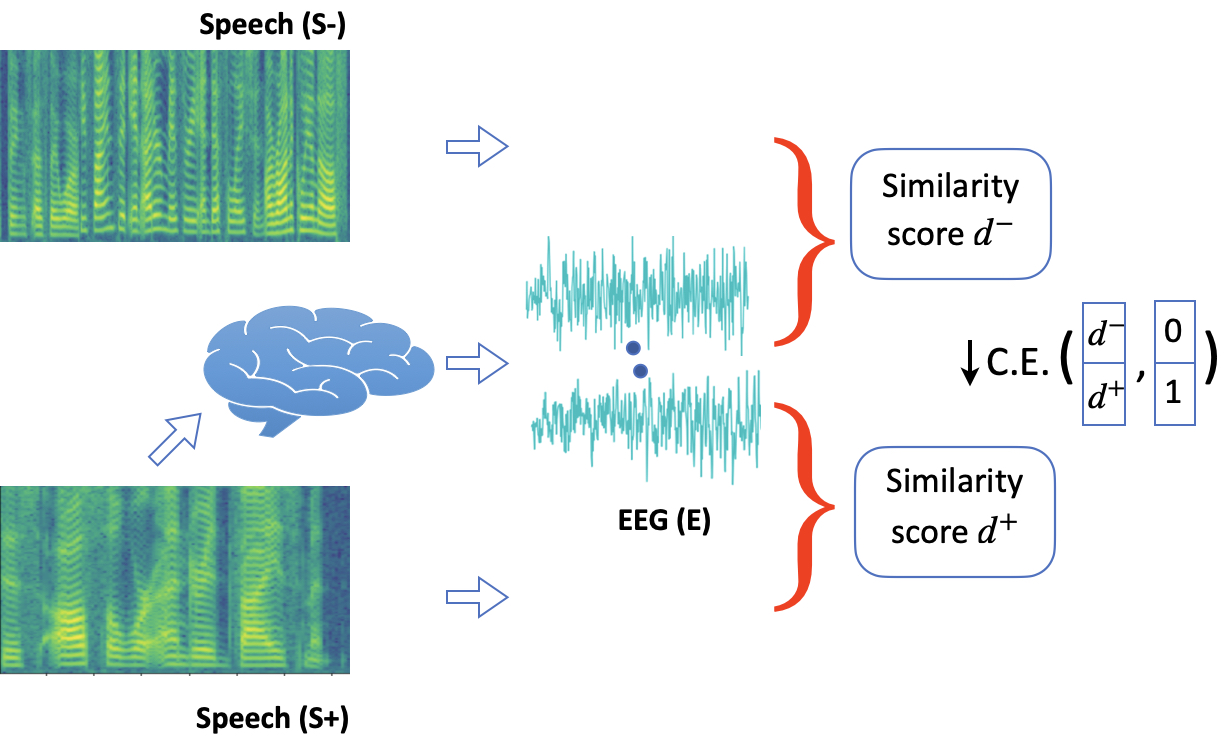}
  \caption{\textbf{Match-mismatch classification task:} It is a binary classification paradigm associating the EEG and speech segments. The EEG segment ($\textbf{E}$) and corresponding stimulus sentence ($\textbf{S}+$) form the positive pair while the same EEG and another (unrelated) sentence ($\textbf{S}-$) form the negative pair. The similarity score computation is achieved using the model depicted in Figure~\ref{fig:speech_nw}.  Here, C.E. denotes the cross entropy loss.}
  \label{fig:mmtask}
\end{figure}
\section{Methods} 

\subsection{Dataset}
We experiment with a publicly available speech-EEG data set\footnote{https://doi.org/10.5061/dryad.070jc} released by Broderick et al. \cite{broderick2018electrophysiological}. It contains electroencephalographic (EEG) data recorded from $19$ subjects as they listened to the narrative speech. The subjects listened to a professional audio-book narration of a well-known work of fiction read by a single male speaker. The data consists of $20$ trials of roughly the same length, with each trial containing $180$s of audio. The trials preserved the chronology of the storyline without repetitions or breaks.  The sentence start and end time, and the word-level segmentation  of the  speech recordings are provided. The word segmentation is obtained using a speech recognition-based aligner \cite{gorman2011prosodylab}. The EEG data were acquired using the $128$-channel BioSemi system at a sampling rate of $512$Hz, while the audio data was played at $16$kHz. Overall, the speech-EEG data amounted to a duration of $19$ hours. 

\subsection{EEG Preprocessing}
The CNSP Workshop 2021 guidelines\footnote{https://cnspworkshop.net/resources.html} served as the basis for the EEG pre-processing pipeline. It is implemented using the  EEGLAB software~\cite{delorme2004eeglab}.The EEG signal is band-pass filtered between $0.5$-$32$ Hz. Then it is down-sampled to $64$Hz. After removing noisy channels  (determined using the channel level statistics), the EEG channels are re-referenced to the mastoids. The data from each channel is also  normalized by computing the z-score. The EEG pre-processing code and the codes used for further analysis discussed in this paper publicly available\footnote{https://github.com/iiscleap/EEGspeech-MatchMismatch}.

\subsection{Acoustic Feature - Mel Spectrogram}
The mel spectrogram of the speech signal is used as the stimulus feature.  The mel spectrogram is computed for each sentence. A mel filter bank with $28$ filters distributed in the mel-scale ranging from $0$-$8$kHz frequency is used. The input audio is pre-emphasized with a factor of $0.97$ before windowing. In order to obtain speech features at a sampling frequency of $64$Hz, the spectrogram computation uses a Hamming window function of the width 
$31.25$ms with half overlap.   
\begin{figure*}[t!]
  \centering
  \includegraphics[width=0.85\textwidth]{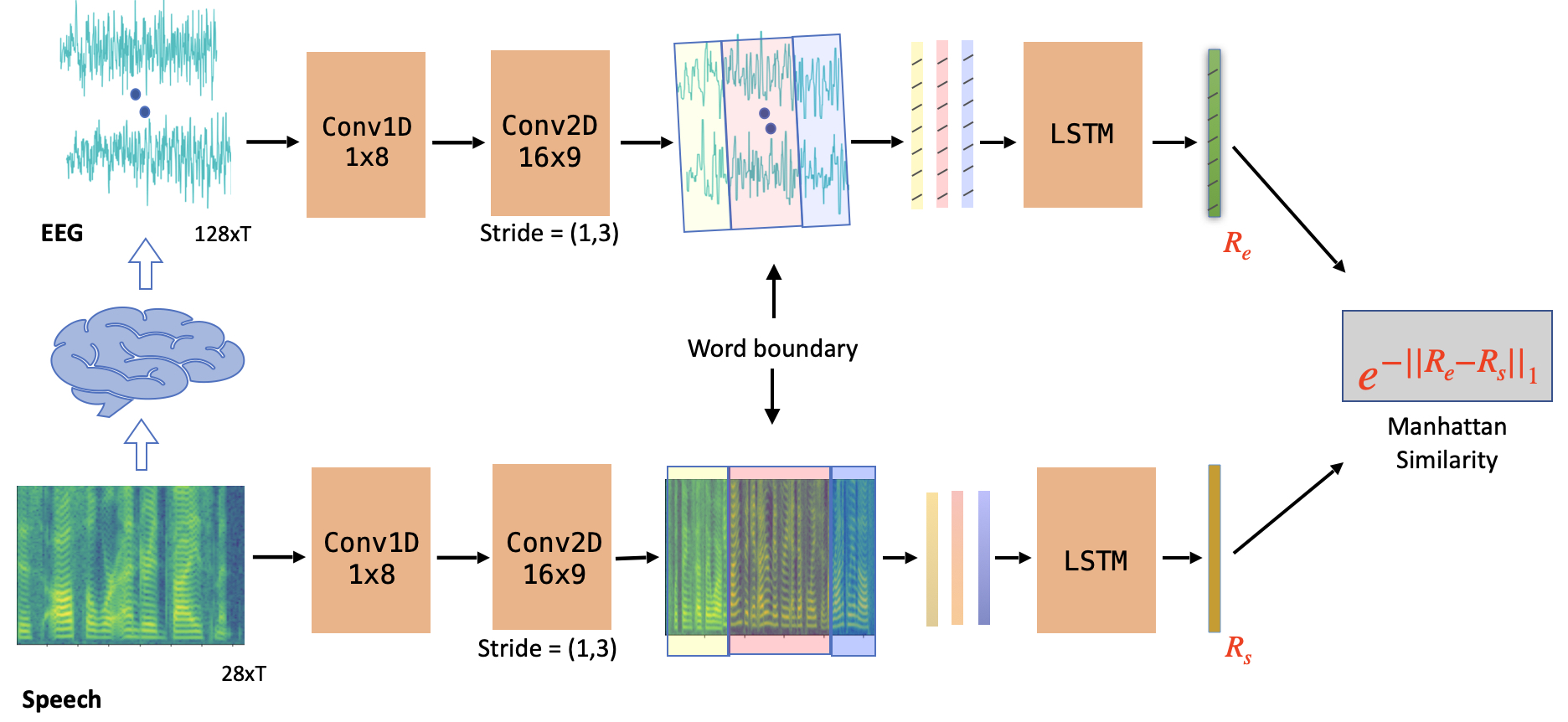}
  \caption{Proposed model for match mismatch task on speech EEG  data. The model training paradigm is outlined in Figure~\ref{fig:mmtask}. } 
  \label{fig:speech_nw}
\end{figure*}
\subsection{Match-mismatch classification task}
The accuracy of a match-mismatch classification task is employed in this study as a measure of the neural tracking of speech. \figref{fig:mmtask} illustrates this paradigm in detail. The classification model is trained to relate the speech segment to its corresponding EEG response. In this study, the segment is chosen to be a sentence. We also compare with prior works \cite{monesi2020lstm,jalilpour2021extracting}, which perform this task at the sentence level.  The time-synchronized stimulus of the EEG response segment is the matched speech. Another sentence from the same trial of data collection is chosen as the mismatched speech. Selecting mismatched samples from the same trial makes the classification task challenging enough to encourage the model to learn the stimulus-response relationships. This sampling approach also avoids the chances of memorizing the speech features along with its label. The matched EEG response for these speech sentences is also included in the mini-batch training to ensure that memorisation is disallowed. 

\subsection{Model architecture}
We employed different modelling paradigms to analyze the encoding of acoustic and semantic features in EEG signals. 

\subsubsection{Baseline Model}
Recently, Monesi et al. \cite{monesi2020lstm} showed that convolutional neural network (CNN) and long short-term memory (LSTM) based architectures outperform linear models for modelling the relationship between EEG and speech. This work employed a match mismatch classification task on fixed duration windows of speech and their corresponding EEG data. The work also demonstrated that mel spectrogram features of the speech stimulus provide the best neural tracking performance compared to other representations like speech envelope, word embedding, voice activity and phoneme identity \cite{jalilpour2021extracting}. They have performed the match mismatch task of $5$s duration segments with $90$\% overlap between successive frames. The prior works \cite{monesi2020lstm,jalilpour2021extracting} use an angular distance between EEG and speech representations, average pooling over time, and a sigmoid operation. The model is trained with binary cross entropy loss \cite{jalilpour2021extracting}. 
We use this approach as the baseline setup for the proposed framework.  

\subsubsection{Proposed match mismatch Model}
The speech signal representation $\mathcal{\mathbf{S}}$ is the mel-spectrogram of dimension $28 \times T$, where $T$ denotes the duration of a speech sentence at $64$Hz. Similarly, the EEG data for the same sentence is denoted as $\mathcal{\mathbf{E}}$, and it is of dimension $128 \times T$. 

Both the speech and the EEG features are processed through a parallel neural pipeline, as depicted in \figref{fig:speech_nw},  without any weight sharing. This sub-network consists of a series of  convolutional layers and LSTM layers.
The convolutional layers implement $1$-D and $2$-D convolutions with $1 \times 8$ and $16 \times 9$ kernel sizes, respectively. The 1-D and 2-D layers have $8$ and $16$ kernels, respectively.  Further, the $2$-D CNN layers also introduce a stride of $(1,3)$ to further down-sample the feature maps. 
\begin{table}[t!]
\centering
\caption{Match mismatch classification accuracy of baseline model \cite{monesi2020lstm} for fixed duration sequences. The step size between adjacent frames is $0.5$s.}
\label{tab:fixed}
\begin{tabular}{c|c}
\hline
\textbf{\begin{tabular}[c]{@{}c@{}}Frame \\ Width (sec.)\end{tabular}} & \multicolumn{1}{c}{\textbf{\begin{tabular}[c]{@{}c@{}}Test\\ Accuracy (\%)\end{tabular}}} \\ \hline
1 & 62.21 \\
3 & 72.41 \\
5 & 76.12 \\
 \hline
\end{tabular}
\vspace{-3ex}
\end{table}

The word boundary information available in the dataset is converted to the equivalent sampling rate (both EEG and audio representations at $\frac{64}{3}$ Hz). The audio and EEG feature maps are average pooled at the word level using the word boundary information.   As a result, for a given sentence, the EEG and speech branches generate vector representations sampled at the word level. An  LSTM layer models the inter-word context from these  representations. This layer is included in both the stimulus (speech) and response (EEG) pathways.  
The last hidden state of the LSTM layer, of dimension $32$, is used as the embedding for the stimulus/response, denoted as $R_s$/$R_e$ respectively.  

We propose the Manhattan distance between the stimulus and response embeddings. The similarity score is computed as,
\begin{equation}\label{eq:eq1}
 d(\mathcal{\mathbf{E}}, \mathcal{\mathbf{S}}) = \exp (- || R_e - R_s ||_1)    
\end{equation}
The similarity score for the matched pair ($\mathcal{\mathbf{E}}, \mathcal{\mathbf{S}^+}$) and mismatched pair ($\mathcal{\mathbf{E}}, \mathcal{\mathbf{S}^-}$) are computed. The model, with a dropout factor of $0.2$, is trained using a binary cross-entropy loss, with [$d(\mathcal{\mathbf{E}}, \mathcal{\mathbf{S}^+})$, $d(\mathcal{\mathbf{E}}, \mathcal{\mathbf{S}^-})$] mapped to [$1$, $0$] targets. 

\begin{table}[]
\centering
\caption{The match-mismatch classification accuracy of speech stimulus and its EEG responses in sentence level for baseline \cite{monesi2020lstm} and the proposed model.  Here, a random speech sentence was chosen as the mismatch sample for each EEG sentence.}
\label{tab:baseline_comparison}
\begin{tabular}{c|cccc}
\hline
       & & \multicolumn{3}{|c}{\textbf{Proposed Model}}                                                           \\ \cline{3-5} 
\multirow{-2}{*}{\textbf{Test Set}} &
  \multicolumn{1}{c|}{\multirow{-2}{*}{\textbf{\begin{tabular}[c]{@{}c@{}}Baseline \\ Model\end{tabular}}}} &
  \multicolumn{1}{c}{\textbf{Cos.}} &
  \multicolumn{1}{c}{\textbf{Euclidean}} &
  \textbf{Manhattan} \\ \hline
Fold 1 & \multicolumn{1}{c|}{65.39} & \multicolumn{1}{c}{88.22} & \multicolumn{1}{c}{93.49} & 94.02    \\
Fold 2 & \multicolumn{1}{c|}{65.32} & \multicolumn{1}{c}{88.73} & \multicolumn{1}{c}{93.68} & 94.00 \\
Fold 3 & \multicolumn{1}{c|}{64.98} & \multicolumn{1}{c}{86.54} & \multicolumn{1}{c}{93.72} & 93.91 \\ \hline
\textbf{Average} &
  \multicolumn{1}{c|}{65.23} &
  \multicolumn{1}{c}{ 87.83} &
  \multicolumn{1}{c}{ 93.63} &
  {\textbf{93.97}} \\ \hline
\end{tabular}
\vspace{-3ex}
\end{table}

\subsubsection{Training and Evaluation Setup} 
The dataset contained recordings from $19$ subjects. All the experiments reported in this work perform subject-independent evaluation (the subjects used in training are not part of the evaluation). Further, we report the average results of $3$-fold validation, with classification accuracy as the metric.  
The experiments are run with a batch size of $32$. The models are  trained using Adam optimizer with a learning rate of $0.001$ and weight decay parameter of $0.0001$. The models are learned with a binary cross-entropy loss.

\section{Results and Discussion}

\subsection{Baseline model on fixed duration segments.}
The baseline implementation for comparison is the work reported in Monesi et al.  \cite{monesi2020lstm}. This architecture is an LSTM model that operates on fixed-duration audio EEG data.  All experiments are run for $20$ epochs of training. The result of the model with fixed duration frames is given in \tabref{tab:fixed}. In order to increase the amount of training data, we also use $90$\% overlap between segments.

\subsection{Baseline model at sentence level}
The baseline model architecture is implemented for fixed-duration segments in training and testing. In order to operate at the sentence level, we have modified the dot product operation as element-wise multiplication followed by an average pooling. This score is passed through the sigmoid function, and the model is learned on sentence-level audio-EEG pairs. For the mismatch condition, a random speech spectrogram is paired with the EEG to generate the score. These results are reported in Table~\ref{tab:baseline_comparison}. 

\subsection{Proposed model with sentence level processing}
\begin{figure}
  \centering
  \includegraphics{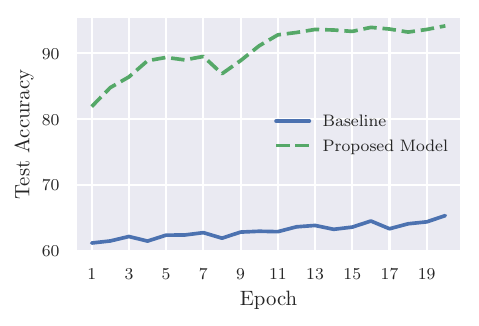}
  \vspace{-0.25in}
  \caption{This figure shows the match-mismatch classification accuracy of the proposed model for test fold-1 as a function of the training epoch for the baseline model and the proposed approach.}
  \label{fig:result_main}
\end{figure}
The results with the proposed model are also reported in Table~\ref{tab:baseline_comparison}. We compare three different similarity scoring approaches, i) Angular (Cosine) similarity, ii) Negative L2 distance (Euclidean) and iii) proposed Manhattan similarity (Eq.~\ref{eq:eq1}). As seen in the results, the Euclidean and Manhattan similarity improves over the cosine similarity.  The proposed EEG-speech match-mismatch classifier model reports an average accuracy of $93.97\%$, which is statistically significantly higher than the baseline model's sentence-level performance (Wilcoxon signed-rank test, $p<1e-4$). The epoch-wise accuracy for test fold-1 is also illustrated in \figref{fig:result_main}.


\subsection{Mismatch sample selection for sentence processing}
\label{sec:mismatch}
\begin{table}
\centering
\caption{Impact of mismatch sample selection strategy on classification accuracy.}
\label{tab:mismatch}
\begin{tabular}{l|r}
\hline
\multicolumn{1}{c|}{\textbf{\begin{tabular}[c]{@{}c@{}}Mismatch Selection \\ Strategy\end{tabular}}} &
  \multicolumn{1}{c}{\textbf{\begin{tabular}[c]{@{}c@{}}Test \\ Accuracy (\%)\end{tabular}}} \\ \hline
Random Sentence &
  93.97 \\ 
Next sentence &
  91.56 \\ \hline
\end{tabular}
\vspace{-3ex}
\end{table}

 Previous match-mismatch EEG-speech studies \cite{de2021auditory,monesi2020lstm} dealt with fixed-duration speech and EEG segments. Cheveigne et al. \cite{de2021auditory} used an unrelated random segment as a mismatched sample, while studies like \cite{monesi2020lstm,accou2020modeling,jalilpour2021extracting} employ a neighbouring segment as the mismatched sample. 
The sampling of the mismatched segments from the same trial ensures that the distribution of the matched and mismatched segments is similar. We explore a similar strategy for  sentence-level analysis by selecting the neighbouring sentence in the same trial as the mismatched sample.
Table~\ref{tab:mismatch} shows how the  mismatch selection strategy affects the classification accuracy.  The average accuracy has a slight degradation when the next sentence is used as the mismatch sample.  
 
\subsection{Importance of accurate word boundaries}
\label{sec:random_word_boundary}
\begin{figure}
  \centering
  \includegraphics{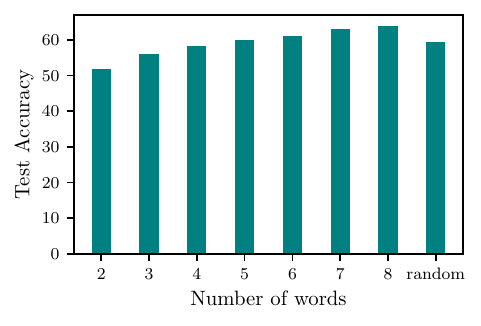}
    \vspace{-0.25in}
  \caption{Average match-mismatch classification accuracy of the proposed model for random word boundaries.}
  \label{fig:result_randomW}
\end{figure}
We conducted several ablation tests to understand the impact of the word boundary information. The model is fed with random word boundaries in the first set of experiments. Each sentence is assumed to contain a fixed number of words and their boundaries are chosen at random.  The results are reported in Figure~\ref{fig:result_randomW}. 
The accuracy improves gradually when the number of word boundaries is increased, even though they are random. The accuracy of the experiment using $8$ words in a sentence is $64\%$, which is significantly lower than the model's performance with accurate boundary information (Wilcoxon signed-rank test, $p<0.0001$).
The final experiment shown in \figref{fig:result_randomW} assumes a random number of words in each sentence with random boundaries, and it provided an accuracy of $60\%$.

In the second set of experiments, we provide accurate word boundary information but skip the word boundary information at every  $n$-th word. These results are reported in Table~\ref{tab:skip-table}. For example, Skip-3 in this table corresponds to removing the word boundary inputs at every $3$-rd entry. The pooling is done with the rest of the available word boundaries for these experiments. As seen in Table~\ref{tab:skip-table}, the results with a higher value of $n$ (of skip-$n$ experiments), approach the setting without any removal (accuracy of $93.97$\%). It is also noteworthy that, even with the Skip-2 setting (word boundary information available for every alternate word), the performance is $82.3$\%, significantly better than the baseline model. 
This study also demonstrates that accurate word boundary information significantly impacts the match mismatch classification, which further illustrates that the EEG signal encodes the word level tracking of speech. 
\begin{table}[]
\centering
\caption{Accuracy (\%) in match mismatch task for varying levels of word-boundary information. Here, $Skip-n$ denotes removing every $n$th word boundary information in the model.}
\label{tab:skip-table}
\begin{tabular}{l|llll}
\hline
\textbf{Test set} & Skip-2     & Skip-3     & Skip-4     & Skip-5     \\ \hline
Fold 1                                                                      & 82.45 & 88.96 & 90.43 & 90.64 \\
Fold 2                                                                      & 81.86 & 88.77 & 90.32 & 90.28 \\
Fold 3                                                                      & 82.60 & 88.79 & 90.30 & 90.01 \\ \hline
\textbf{Average}                                                            & 82.30 & 88.84 & 90.35 & 90.31 \\ \hline
\end{tabular}
\vspace{-3ex}
\end{table}

\section{Conclusions}
In this paper, we have attempted to validate the hypothesis that speech comprehension in the brain is segmented at the word-level in the EEG responses to continuous speech. 
For this task, we developed a deep neural network model consisting of convolutional encoders, word-level aggregators and recurrent layers. A novel loss function for this task based on Manhattan similarity is also proposed. 
The proposed model validated the hypothesis by improving the accuracy of match-mismatch classification of speech and EEG responses at the sentence level. The incorporation of word boundary information yields statistically significant improvements compared to the baseline model, demonstrating the importance of this information in the neural tracking of speech. Moreover, the proposed model handles variable length inputs.   Overall, this model can have potential applications in various domains, including speech recognition, brain-computer interfaces, and cognitive neuroscience. Future research could explore this model's extension to incorporate multi-modal inputs in the form of textual data in addition to the speech spectrogram.


\bibliographystyle{IEEEtran}
\bibliography{mybib}

\end{document}